# That's How We Roll – The NASA K2 Mission Science Products and Their Performance Metrics

Jeffrey E. Van Cleve[1,2], Steve B. Howell[1], Jeffrey C. Smith[1,2], Bruce D. Clarke[1,2], Susan E. Thompson[1,2], Stephen T. Bryson[1], Mikkel N. Lund[3,4], Rasmus Handberg[4], William J. Chaplin[3,4]

## Abstract

NASA's exoplanet Discovery mission *Kepler* was reconstituted as the *K2* mission a year after the failure of the 2nd of *Kepler*'s 4 reaction wheels in May 2013. Fine control of the spacecraft pointing is now accomplished through the use of the two remaining well-functioning reaction wheels and balancing the pressure of sunlight on the solar panels, which constrains *K2* observations to fields in the ecliptic for up to approximately 80 days each. This pseudo-stable mechanism gives typical roll motion in the focal plane of 1.0 pixels peak-to-peak over 6 hours at the edges of the field, two orders of magnitude greater than typical 6 hour pointing errors in the *Kepler* primary mission. Despite these roll errors, the joint performance of the flight system and its modified science data processing pipeline restores much of the photometric precision of the primary mission while viewing a wide variety of targets, thus turning adversity into diversity. We define *K2* performance metrics for data compression and pixel budget available in each campaign; the photometric noise on exoplanet transit and stellar activity time scales; residual correlations in corrected long cadence light curves; and the protection of test sinusoidal signals from overfitting in the systematic error removal process. We find that data compression and noise both increase linearly with radial distance from the center of the field of view, with the data compression proportional to star count as well. At the center, where roll motion is nearly negligible, the limiting 6 hour photometric precision for a quiet 12th magnitude star can be as low as 30 ppm, only 25% higher than that of *Kepler*. This noise performance is achieved without sacrificing signal fidelity; test sinusoids injected into the data are attenuated by less than 10% for signals with periods up 15 days, so that a wide range of stellar rotation and variability signatures are preserved by the *K2* pipeline. At time scales relevant to asteroseismology, light curves derived from *K2* archive calibrated pixels have high-frequency noise amplitude within 40% of that achieved by *Kepler*. The improvements in *K2* operations and science data analysis resulting from 1.5 yr of experience with this new mission concept, and quantified by the metrics in this paper, will support continuation of *K2*'s already high level of scientific productivity in an extended *K2* mission.

1. NASA Ames Research Center, Moffett Field, CA 94035; jeffrey.vancleve@nasa.gov
2. SETI Institute
3. School of Physics and Astronomy, University of Birmingham, Edgbaston, Birmingham, B15 2TT, UK
4. Stellar Astrophysics Centre (SAC), Department of Physics and Astronomy, Aarhus University, Ny Munkegade 120, DK-8000 Aarhus C, Denmark

## 1. Introduction: *K2* is born from *Kepler*

NASA's *Kepler* exoplanet transit Discovery mission (Borucki et al., 2010) launched in 2009 and completed its primary mission of 4 years of nearly continuous high-precision photometry of over 100,000 stars in Cygnus and Lyra, with abundant scientific results in the fields of exoplanets (Mullally et al. 2015), asteroseismology (Chaplin et al. 2011), and stellar rotation (Meibom et al. 2011) and variability.  A key result was that the intrinsic photometric noise of typical solar-type stars was in fact somewhat higher than that of our Sun (Gilliland et al. 2011), and an extended *Kepler* mission was defined to overcome this additional noise by continuing observations of the same celestial field of view (FOV) for an additional 2+ yr.  The loss of 2 of *Kepler*'s 4 reaction wheels by May 2013 ended the *Kepler* spacecraft's ability to carry out its original mission, and over the next year the science and operations teams formulated and tested the *K2* mission concept which would use the otherwise well-functioning spacecraft and photometer, a 0.95-m Schmidt telescope with a 100 Mpix focal plane array (FPA) with 84 output channels (Van Cleve & Caldwell, 2009).  In this paper, we will refer to the mission before the 2$^{nd}$ wheel failure as *Kepler*, and the subsequent science mission as *K2*.  *Kepler* quarterly data sets are denoted as Q[m] for m = 1 to 17, while *K2* campaigns are denoted as C[n] for n = 0 to 7 as of this writing.

The critical insight for the *K2* mission is that the spacecraft is able to maintain three-axis pointing control using the two remaining reaction wheels to control the telescope boresight, while balancing the solar radiation pressure on the solar panels to minimize torques around the roll axis (Figure 1;  Putnam & Wiemer, 2014).  The only way this symmetry can be maintained for an extended period of time is to point the telescope in the ecliptic, in which case observing campaigns of up to 80 days are possible in a window bounded by the times that the Sun is too low over the solar panels to provide power, or too close to the telescope boresight for stray light and telescope safety.  Like balancing a pencil on your finger, this orientation is an unstable equilibrium, and the telescope roll axis must be kept within bounds by thruster firings regularly scheduled on 6 h = 12 Long Cadence (LC) intervals.  In addition, the momentum accumulated on the two remaining reaction wheels in order to hold the boresight position must be dumped every two days.  While both C0 and C1 had a mid-campaign break for data downlink, all subsequent campaigns – with the exception of the customized C9 microlensing campaign -- collect data continuously and hence require a good understanding of why data volume per target is greater for *K2* than *Kepler* (Sections 2.2 and 2.3).

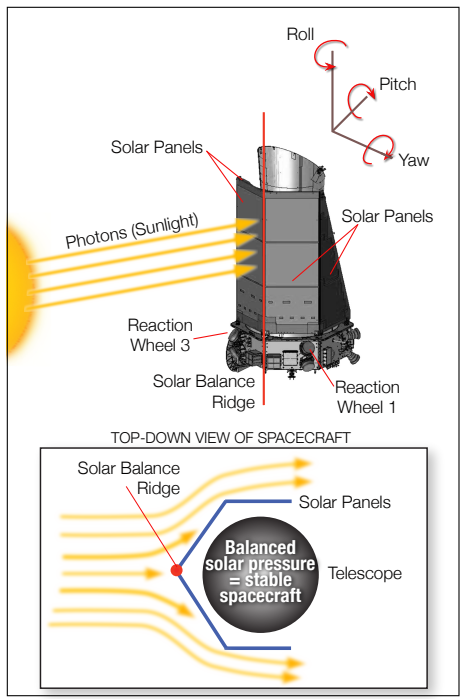

**Figure 1:  *K2* -- how we roll.**  *K2* uses its two remaining reaction wheels to point the telescope to a field on the ecliptic.  Like balancing a pencil on its point, the roll axis is stabilized by the balance of solar

radiation pressure and small thruster firings at 6 h intervals.  See the online edition for a color version of this figure.

With three-axis pointing thus restored, the pointing at the edge of the FOV is within 1.4 pixels or 5.6" of nominal for 95% of a campaign, compared to <0.1 pixels in the original mission. This roll motion impresses a sawtooth waveform with an amplitude of several percent on the uncorrected simple aperture photometry (SAP) light curves (Figure 2) as targets move around in their fixed apertures; the sawtooth amplitude is over two orders of magnitude larger than the photometric precision achieved by *Kepler*.  The attitude error and resulting sawtooth photometric signature is complex (though quasiperiodic) on both long and short timescales: the perturbing torque changes sign between the beginning and end of a campaign, with a low torque period in the middle during which some scheduled thruster firings are skipped, and there is fine structure on a roughly 3 min time scale in an active thruster firing window during which a series of thruster firings are executed as the attitude control system homes in on the desired stable fine pointing attitude. In the face of such large and complex systematic errors it was necessary to prove that *K2* was still capable of photometric precision superior to that of the best ground-based photometry to justify the mission.  Howell et al. (2014), following Gilliland et al. (2011), used a simple Savitsky-Golay (SG) filter to clean the data and showed that the 6-hr photometric precision for an uncrowded $12^{th}$ magnitude dwarf star increased from ~20 to ~80 ppm, still almost a factor of 2 better than the best state-of-the-art ground-based photometry (Everett and Howell, 2001; Nascimbeni et al. 2013).  In Section 4.2, we discuss how the SG filter overestimates noise for exoplanet detection, and how noise varies across the FPA, and show that the actual performance is significantly better than the early estimate by Howell et al. (2014).

Several fundamental axioms of *Kepler*'s concept of operations and design (Van Cleve & Caldwell, 2009) were violated by *K2*, and creatively worked around by science and operations teams (Peterson et al., 2015) for the initial *K2* concept demonstration, and over the subsequent year of experience with operations and data analysis.  First, the FOV changes every 80 days instead of being fixed for the entire mission, requiring a new round of target selection, target pixel definition, and attitude change maneuvers. Second, the *Kepler* data volume and downlink budget was based on the assumption that the time variation of the signal on a pixel would typically be much smaller than the mean signal, and hence that the data could be highly compressed.  The roll motion in *K2* makes sequential pixels more unlike each other than in *Kepler*, and hence they require almost twice as many bits per pixel to encode.  Third, in *Kepler* the allocation of target pixels assumed that the position of a star on the FPA could be predicted and maintained to better than 0.1 pixels, so that only the pixels needed for extracting light curves needed to be assigned. In K2, "halos" of extra pixels must be assigned to collect the right ones even when pointing error is at its maximum, which at the edge of the FOV can be 2.0 pixels.  These extra halos typically increase the number of pixels assigned by 2-4x.  The net result of worse compression and increased halo pixels is that the number of targets available decreased from 170,000 in *Kepler* to 15,000 -- 25,000 in *K2,* depending on field crowding.  The actual number of targets allocated may be less in some campaigns if large numbers of pixels are dedicated to bright or moving targets, or tiling large areas of the sky. In Sections 2.2 and 2.3, we discuss how halo allocation and data compression vary with position on the focal plane and star count, and how they have improved since Howell et al. (2014).

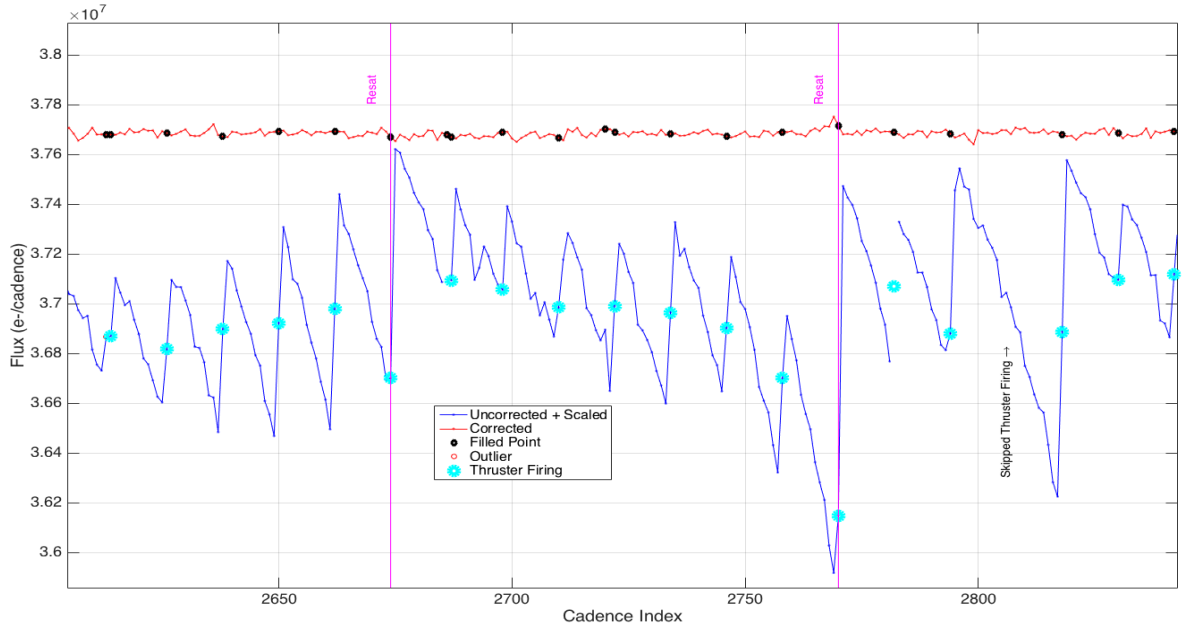

**Figure 2: Example sawtooth pattern impressed on light curve by *K2* roll motion and thruster firings.** The uncorrected SAP light curve has been scaled to the expected mean flux of the target alone. Near cadence index 2800 is an example of a scheduled thruster firing which is skipped so that 12 h passes between firings. Lines marked 'Resat' indicate the reaction wheel resaturations which occur every 2 days to control yaw and pitch. While the RMS amplitude of the sawtooth in this example is ~1.0%, the modified *K2* data analysis pipeline (corrected light curve) reduces the effective noise for transit detection by almost two orders of magnitude, to 127 ppm for a star with $K_p$ = 14.4. See the online edition for a color version of this figure.

The *K2* mission also changed several basic attributes of the data, which required the rebuilding of the *Kepler* data analysis pipeline. In order to select the fixed subset of pixels used for light curve extraction from the set of pixels collected, image motion and signal-to-noise have to be examined over the course of a campaign and an overall optimal aperture defined (Smith et al., in prep). Correcting light curves for systematic error in the Pre-search Data Conditioning (PDC) module of the pipeline also became more difficult in two ways: there are too few targets per output channel of the focal plane to construct a systematic error basis set and a local population of Bayesian priors (Stumpe et al., 2012 and Smith et al., 2012), and the sawtooth (Figure 2) is both large and on the same time scale as exoplanet transits, confounding the separation of systematic errors by timescale which had proved fruitful in the *Kepler* data (Stumpe et al., 2014). The results in Section 4 show the noise performance of corrected light curves after *K2* pipeline changes.

Finally, the *K2* project has responded to K2's capabilities by changing the science selection process. *Kepler* emphasized measuring the probability of Earth-sized planets in the habitable zone (HZ) of solar-type (late K and early G dwarfs), closely supported by asteroseismology studies of parent stars. *K2* science still includes dwarf star exoplanets (Crossfield et al. 2015; Vanderburg et al. 2015; Foreman-Mackey et al. 2015; Petigura et al. 2015), but now has increased emphasis on other topics such as variable and eclipsing binary stars (Armstrong et al. 2015), pulsations across the HR diagram such as subgiants (Chaplin et al. 2015), and AGN variability (Shaya et al. 2015). *K2* has wide community involvement in field of view (FOV) selection, and all targets are selected within each FOV through an open proposal process run by NASA and the *K2* Guest Observer (GO) office. All *K2* data are publically available, without proprietary period, after pipeline processing. While GOs are observing targets which may host exoplanets, the *K2* mission itself is *not* doing its own transiting planet searches, leaving that up to the creativity of the community. The ultimate result is that the *Kepler* project has turned adversity into diversity for science topics and data analysis methods.

## 2. Target Management

### 2.1 EPIC

The Ecliptic Plane Input Catalog (EPIC, Huber et al., 2015) plays the same role for *K2* that the *Kepler* Input Catalog (KIC, Brown et al. 2011) played for *Kepler* target selection. The primary purpose of the catalog is to provide celestial positions and *Kepler* 450-825 nm bandpass magnitudes ($K_p$) for each GO-selected target, in order to select target pixels for downlink. A secondary goal is to provide estimates of stellar properties to facilitate target selection. EPIC is hosted at MAST (http://archive.stsci.edu/k2) and should be used for selecting targets whenever possible. EPIC is updated for future *K2* campaigns as their fields of view are finalized and the associated target management is completed.

EPIC parameters are produced by federating existing multi-band catalogs and by calculating color corrections for the *Kepler* bandpass. While EPIC is complete to *Kepler* magnitude $K_p$ < 17 and typically accurate to ~0.1 mag (Huber et al, 2015) unless otherwise noted in the campaign Data Release Notes, observers can propose targets that are fainter than the typical completeness limits and/or are not presently included in the Ecliptic Plane Input Catalog. Such targets will be added to the catalog in future deliveries if selected for observation.

While the EPIC catalog lacks sky survey data in the custom narrow Mgb 510 nm filter, which was used in the KIC because of its sensitivity to surface gravity and to metallicity, Huber et al. (2015) used the KIC as a training set to classify 85% of the full *K2* target sample for C1-C7 using colors, proper motions, spectroscopy, parallaxes, and galactic population synthesis models, with typical uncertainties for G-type stars of ~ 3% in Teff , ~ 0.3 dex in logg, ~40% in radius, ~10% in mass and ~40% in distance.  Since the errors in the EPIC logg are now comparable to those of the KIC, ~95% confidence in distinguishing dwarfs and giants is now restored for already-selected *K2* targets.  However, because of the vastly greater number of EPIC sources over the first 13 distinct *K2* FOVs (>2.7x10$^7$), the existing sky survey data and the analysis tools of Huber et al. (2015) will not be providing high-quality stellar property estimation for the entire EPIC catalog in advance of target selection, as was possible with the KIC, for the foreseeable future.

### 2.2 Aperture Size

The Target Aperture Definition (TAD) module of the *Kepler* Pipeline (Bryson et al. 2010) selects the pixels for each target which will be downlinked, a subset of which will be used to form the photometric aperture during light curve extraction (Section 3.2.2).  TAD works with a pixel budget determined by the size of the Solid-State Recorder (SSR) and estimated data compression.  As in the *Kepler* mission, TAD works by generating a synthetic image from the Pixel Response Function (PRF) and a star catalog (Section 2.1) and generating a list of pixels which maximizes signal-to-noise (SNR).  This pixel list was augmented by a "halo" of pixels to guarantee that the SNR-maximizing pixels were collected in the presence of absolute pointing errors, and an additional column of pixels to characterize electronic undershoot upstream of the target (Van Cleve & Caldwell 2009). Unlike *Kepler*, for which the only appreciable image motion was at most 0.6 pixels of differential velocity aberration (DVA) over a ~90 day period, *K2* apertures must be oversized by an additional halo in the FOV center and two halos at the edge in order to assure that all the pixels in the union of per-cadence optimal apertures (Section 3.2.2) are included.  This oversizing has a considerable impact on the number of pixels per target; for *Kepler*, the smallest mask (for one pixel optimal aperture) was thus 4x3 = 12 pixels, increasing in *K2* to 6x5 = 30 for inner FOV (2 total halos) and 8x7 = 56 pix for outer FOV (3 total halos). Early campaigns (C0-C4) assigned even more halo pixels, given the uncertainties in performance and the ~6 month time lag between pixel assignment and data analysis.

C7 is a good benchmark for *K2* targeting performance, since the lessons of C0-C4 have been applied, and the average star count over the FOV in C7 is the closest to that of *Kepler* of any *K2* campaign to date. Figure 3 shows the C7 mask sizes, illustrating the limiting, small-mask (faint-source) values, the dependence on *Kepler* magnitude $K_p$ and radial distance $r_{FOV}$ from the FOV center, and the high cost in pixels of bright targets.  A total of 13,483 LC stellar targets were scheduled in C7, about 10% of the number of LC targets towards the end of *Kepler*, though many of the C7 pixels were dedicated to bright or moving targets, or to tiling a cluster. For a comparison to *Kepler*, Fig 6 of Bryson et al. (2010) shows a mode of 80 pixels per mask at 10th mag while C7 modal value at 10th mag is around 180.  At 15th mag, the pixel count goes from a unimodal 22 pixels per mask for *Kepler* to a noticeably bimodal distribution with maxima of 50 and 80 pixels per mask for the inner and outer parts of the FOV, respectively, in *K2* C7.  Averaged over the target set, there are ~4x the number of pixels per target and up to 2x the number of bits per pixel (Section 2.3) to fit into the same SSR budget.

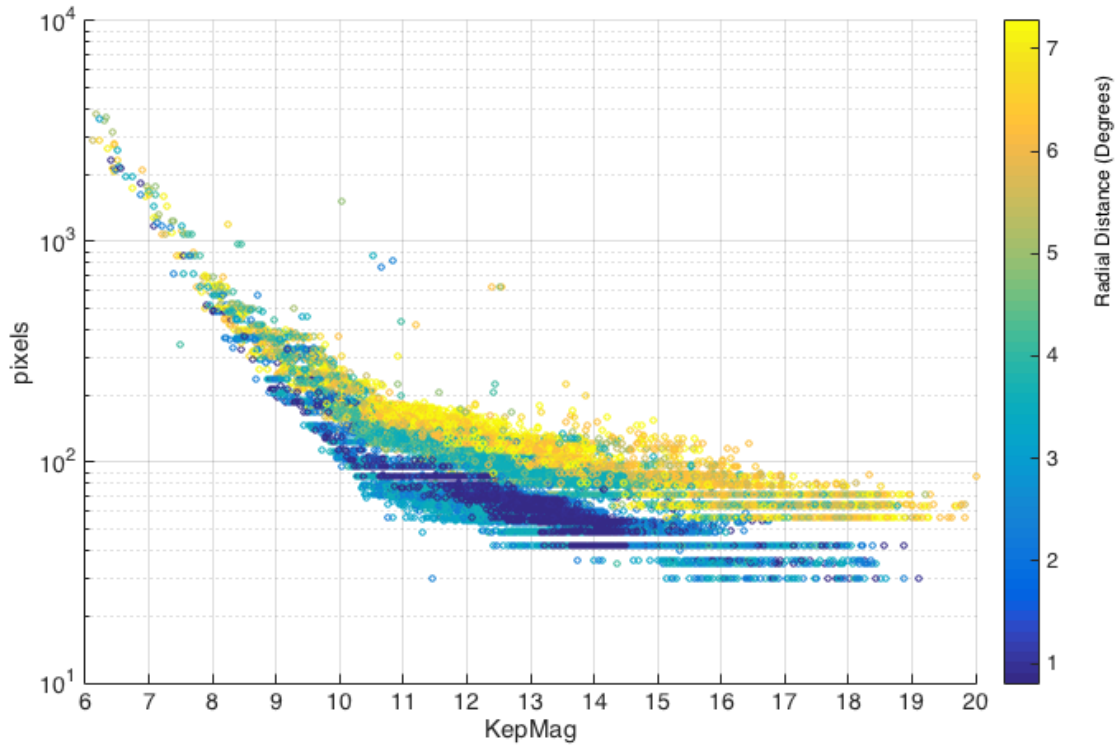

**Figure 3:  Number of pixels in target mask vs. KepMag and radial distance from FOV center for C7, which has a star count closest to that of the *Kepler* primary mission.** For faint stars, masks near the center of the FOV, where roll is smaller, require only two halos instead of three to assure capture of the required pixels.  Compare to Fig. 6 of Bryson et al. (2010).

## 2.3   Data Compression

Because of roll, sequential images are not as much like each other as in the past; they contain more information and hence are harder to compress.  The problem is compounded by the need to add additional halos to capture image motion.  The result is that science may be cut short because the on-board SSR fills up and no more data can be collected for that campaign, or some scientifically worthy programs are downsized to ensure that the top targets get a full campaign's worth of data.  Accurate prediction of the compression $B$ in bits per pixel for a given campaign is important to maximize the science return without overfilling the recorder and subsequent data loss.  The *Kepler* $B$ was between 4.6 and 5.4 bits per pixel without significant dependence on FOV location or star count.  *K2* $B$ has a typical mean value of 8 bits per pixel, and varies significantly with both $r_{FOV}$ and with star count $n_*$ (Figure 4), with values as high as 12 bits per pixel in channels near the edges of the FOV which also have high star counts. We have done channelwise linear regression and prediction of $B$ as a function of star count and $r_{FOV}$ for internal planning purposes.  This model may be useful to observers in future campaigns if the GO office begins to budget observations by bits rather than pixels, and will be made available to the GO community at that time if it remains successful at predicting $B$.  Then if targets of equal scientific merit may be found in the center of the FOV or regions of low star count, they will cost fewer bits to achieve the goals of the proposal

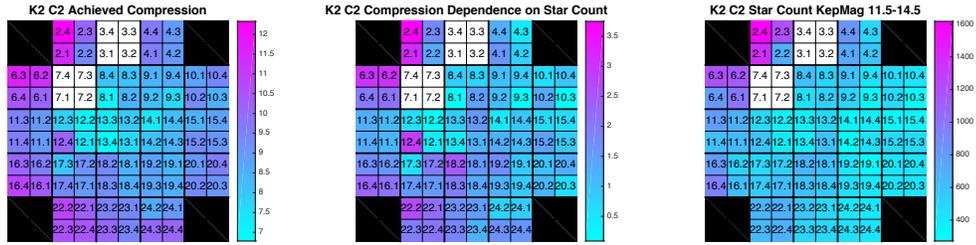

**Figure 4: Example data compression in bits/pixels from campaign 2, shows both radial and star count terms.**  Left:  achieved compression.  Middle:  Compression with best-fit constant and linear terms removed.  Right:  Number of 11.5-14.5 magnitude stars per square degree.  Channels are labeled by module and output.  White indicates nonfunctioning modules.

## 3. Archive Products

### *3.1  K2 Data Products*

All of the *K2* data is available at the MAST http://archive.stsci.edu/k2/ , and *K2* documentation at the *Kepler* & *K2* Science page http://keplerscience.arc.nasa.gov/index.html.  For those familiar with the *Kepler* data products, the *K2* data products are remarkably similar, though Short Cadence (SC) light curves and Reverse Clock (RC) are not available, and LC light curves are not available yet for C0-C2.  The common features of *Kepler* and *K2* data products are described in the *Kepler* Archive Manual (KAM) available from MAST.  See the *K2* Data Release Notes for each Campaign for general target information and performance metrics, and the Pipeline Release Notes for algorithm updates, availability of different types of data, and minor changes to FITS headers.

The *K2* project is not delivering SC light curves, only SC target pixel files and collateral data. However, all the information created by running the pipeline modules that create the light curve, are available to populate the appropriate columns of the target pixel files. For instance, the per-pixel background is available and those items that depend on calculating the motion polynomial (e.g. the per cadence position correction columns) are available.

Targets that are not in the EPIC, or require a specially selected mask, are called custom apertures.  Like *Kepler*, these targets are cataloged using the Custom Aperture number instead of an EPIC number.  Custom aperture numbers range from 200000811--201100000.  For *K2* the custom apertures are frequently used when observing large patches of sky covering an open cluster.  They are also used to observe objects moving across the field of view, such as planets, trans-neptunian objects, or asteroids.  In both cases, the entire section of sky is split between many custom apertures and thus many different data files. The light curve files, while present for these objects, have no real meaning.  To find the correct custom apertures for a particular target at the MAST, search by Investigation Id, or by using the "Object Type" pull down menu on the *K2* Data search page (https://archive.stsci.edu/k2/data_search/search.php).

The astronomical community has been one of the greatest resources for K2. Several groups have calculated their own light curves and made them available through the High Level Science Products at MAST. Currently these products include the detrended light curves from Vanderburg and Johnson (2014) and the light curves and variability catalog of Armstrong et al. (2015).  For some scientific goals, users may find these light curves superior to those produced by the *K2* Project.

### *3.2  K2 Data Processing*
#### 3.2.1  Target Pixels

Target pixel calibration for *K2* is performed using the *Kepler* mission Pipeline module CAL (Quintana et al. 2010) without algorithm modification. What has changed for *K2* is the availability of data, the pipeline software configuration and scene dependent maps of saturated rows and columns:

Availability. Some types of pixels are permanently or temporarily unavailable in K2:

1) Reverse-clock (RC) pixels are no longer collected, in order to save time for science data collection. This limits the ability of the pipeline to dynamically correct certain types of image artifacts (see below).
2) Although very unlikely, the downlinked *K2* data may contain cadences where some but not all of the collateral pixels are gapped.
3) Exported target pixels for C0-C2 do not include background subtraction or estimates since these require running the Photometric Analysis (PA) module to generate light curves, which was not done in these early campaigns. Users will therefore have to do their own background estimates.

Dynamic Black Correction. Reverse-clocked data is used in the *Kepler* mission processing to dynamically correct the 2D black for temporally and thermally varying crosstalk signals from the fine guidance sensors (FGS crosstalk). This is done on a per cadence basis using the *Kepler* pipeline module Dynablack (Kolodziejczak et al. 2010). Since reversed clocked data is not available in *K2,* a *K2* version of Dynablack has been developed which uses only LC collateral data and Artifact Removal Pixels (see Kolodziejczak et al. 2010 for ARPs). The effectiveness of *K2* Dynablack is still being evaluated, and it has not been used for archive data products yet. The Dynablack-calculated RB_FLAG and RB_LEVEL in the collateral and target pixel files may be ignored until such future time as *K2* Dynablack is adopted.

<u>Scene-dependent saturated rows and column maps</u>. Just as in *Kepler* processing, *K2* pixel calibration requires knowledge of which columns contain saturated charge spill (bleed) into the masked and virtual smear collateral regions. A map of bleeding columns is prepared for each *K2* campaign based on a combination of FFI and Long Cadence (LC) data. Rows which contain bright stars near the trailing black region are also identified from FFI data so that they may be excluded from fitting the trailing black row dependent model. A map of these scene dependent rows is prepared for each *K2* campaign. In *Kepler* processing there were a couple of channels where the bleeding columns (near the trailing black region) and/or scene dependent rows made fitting the trailing black pixels to the row dependent model problematic for certain quarters. In these cases, since any particular CCD would observe the same scene every four quarters, a table of model fit coefficients was developed as a proxy for the model fit in the affected components during the affected quarters. No such coefficient overrides are available for *K2* processing. Users might see an unexpected row dependence in the target pixel values at levels on the order of the calibrated trailing black pixels.

The noise and residual bias performance of the black subtraction in *K2* is the same as that of *Kepler*, indicating that electronic noise induced by component aging and the larger temperature swings of *K2* are not a significant contribution to *K2* noise compared to roll-induced photometric noise.

### 3.2.2   Long Cadence Light Curves

For the early *K2* campaigns C0-C2, MAST has only calibrated target pixel files, so as not to delay pixel data release until new algorithms could be developed for light curve extraction and systematic error correction. Beginning with C3, Pipeline *K2* LC light curves are now available at MAST, and C0-C2 will be reprocessed to produce light curves in the near future.

For light curve extraction in the Photometric Analysis (PA) module, image motion is mitigated by calculating the optimal aperture on a per-cadence basis, instead of an averaged optimal aperture over each quarter as in *Kepler*. The aperture definition process was itself improved by replacing *Kepler*'s pure pixel response function (PRF) and catalog modeling with a data-driven approach which uses the PRF to estimate signal, but the actual pixel data to estimate noise. Then apertures are calculated by accreting pixels to maximize SNR estimated with the same algorithm used to calculate the transit noise metric CDPP (Section 4.2). The 95% union of all such individual per cadence optimal apertures forms a fixed aperture which reduces sensitivity to the large image motion while also minimizing background contamination.

For systematic error correction in the Pre-Search Data Conditioning (PDC) module, several changes had to be made to the highly effective algorithms which were developed for *Kepler*, centered around a Bayesian Maximum a Posteriori (MAP) algorithm (Stumpe et al. 2012, Smith et al. 2012, Stumpe et al. 2014):

1. Since there are typically <20% of the number of targets per channel in *K2* than in *Kepler* (Section 2.2), channels are aggregated into symmetrical regions in order to generate a set of basis vectors spanning the function space of systematic trends in the data using Singular Value Decomposition, or SVD. Since the noise properties of the 5-12 channels so combined can be different, this aggregation may degrade PDC's ability to parsimoniously model systematic errors due to detector properties.
2. Stellar variability is calculated after coarse MAP so that the sawtooth does not dominate the calculated stellar variability.
3. Multi-scale MAP (Stump et al. 2014) is no longer used.
4. 12 basis vectors are used, by default. The *Kepler* primary mission used 8, by default. The systematic trends are now stronger and dominated by the sawtooth. More basis vectors can therefore be safely used, and are also necessary to successfully remove the sawtooth.

5. Residual Sawtooth removal. Sawtooth removal is not always complete. Some sawtooth "teeth" will remain after applying the basis vectors. Sometimes a large negative tooth is added due to poor correlation between the basis vectors and the individual teeth. To remove these artifacts, an exponential filter is fit to the light curve right before each thruster firing. If a large response is detected then a residual sawtooth is identified and the cadences with a large response are gapped. Typically, no more than a couple of sawtooth teeth are partially gapped due to this process.

The most significant resulting change to light curve correction in *K2* is that the Bayesian prior is used only roughly 20% of the time.  For most targets a simple robust least-squares (LS) fit to the basis vectors is best, as determined by the "goodness metric" in PDC (Stumpe et al., 2012).   One way to understand this is that even in *Kepler*, quiet stars were usually fit with LS instead of the prior fit as discussed in Smith et al. (2012), since even small errors in the prior are larger than the true signal.  With most targets in *K2* being astrophysically "quiet" compared to the much larger systematic errors seen in *K2*, the majority of *K2* targets wind up being treated like quiet stars in *Kepler* and corrected with LS fits.  Despite this changeover from Bayesian to LS fitting for most targets, signal fidelity in *K2* is preserved for time scales of less than 15 days as discussed in more detail in Section 6.

While PDC is effective at removing up to 99% of the sawtooth, a residual signature is often visible above the stochastic noise (Figure 2).  While manual inspection can see the problem, developing an algorithm that quantifies this residue as a goodness metric for use internal to PDC in order to more fully remove the sawtooth is a challenge. Passing the corrected lightcurves through the same Savitsky-Golay filter as the CDPP proxy (Section 4.2 ) to enhance contrast shows that the residual sawtooth is diffused from the thruster firing cadences; the running difference of a corrected light curve is not a strict delta function + weakly varying real signal as for a sawtooth, and it's not strictly periodic enough to find by phase-folding. It is also hard to measure and clean spectrally, as was also noted by Lund et al. (2015).  Time-domain division of data into the three roll torque sign intervals (see above) and using our knowledge of when thruster firings occurred will be explored for mitigation in future work.

## 4. Instrument Noise in Long Cadence Light Curves

### 4.1 Noise Metric Philosophy

The *Kepler* Pipeline uses a formal, wavelet-based algorithm to calculate the effective signal-to-noise of the specific waveform of transits of various durations (Jenkins et al. 2010); the effective noise in this detection process is referred to as Combined Differential Photometric Precision (CDPP) on the transit time scale. The performance benchmark for *Kepler* was 6 hr CDPP on a 12$^{th}$ magnitude dwarf star, and this metric is still calculated in the *K2* Pipeline and summarized in this paper. However, since the data analysis approach for *K2* is "…letting a hundred flowers bloom and a hundred schools of thought contend … to promote the … progress of science…" (Mao, 1957), we define noise metrics which are easy to compute and explain (following Gilliland et al. 2011), so that various approaches to *K2* data can be compared without making assumptions about signal waveform. Following Gilliland et al. (2015) we also consider noise metrics on both much longer and much shorter time scales to assess performance for nontransit astrophysical observations. For each metric, we use EPIC (Section 2.1) to select dwarfs using log g > 4, to reduce contamination of the noise metrics by stellar variability (i.e., red giants). Our notational schema is to denote a noise metric by a short string describing the filter applied to the corrected light curve, with a subscript indicating the broadband timescale in hours sampled by the metric. The nth percentile of a distribution of noise is $P_n$(noise). So in the example of CDPP, we will write $CDPP_6$ as the noise of an individual target, and $P_{10}CDPP_6$ for the 10$^{th}$ percentile of a collection of targets, possibly selected by magnitude bin and output channel. Highlighted results will be for 12$^{th}$ magnitude dwarfs, unless explicitly stated otherwise. To compare *K2* to *Kepler*, we choose *Kepler* Q12, which was the noisiest *Kepler* quarter because of a coronal mass ejection, to illustrate how good *Kepler* was at its "worst." It is important to note that the difference between the noisiest and quietest *Kepler* quarters is only 5%.

### 4.2 Transit Timescale

While the typical duration of a transit varies from a few hours for close-in planets to 16 hours for a Mars-size orbit, the *Kepler* project adopted a 6 h transit as its performance benchmark, and CDPP is calculated for transits of this duration among others (Jenkins et al., 2010). We define a proxy metric for $CDPP_6$ following Gilliland et al. (2011, 2015), Howell et al. (2014), and Handberg & Lund (2015): we replaced 5-$\sigma$ outliers remaining in the LC light curves with Gaussian noise of the same median absolute deviation, then applied a high-pass Savitsky–Golay filter (2.0 d = 97 LCs) followed by binning into 6.0 hr (12 LCs) samples as a proxy for CDPP. The filter response of the SG filter and binning is shown in Figure 14 of Gilliland (2011). The result is normalized by the result for zero-mean, unit variance white Gaussian noise (WGN). We call this metric $SG_6$ to distinguish it from wavelet-based CDPP on the same time scale; it is calculated for each star, and binned by $K_p$ and FPA output channel.

Gilliland et al. (2011) found $SG_6$ to be a good proxy for CDPP, which we also find to be the case when we apply the same algorithm to *Kepler* Q12 data (Figure 7). In *K2,* residual sawtooth signals contribute to $SG_6$ more than to CDPP (Figure 7); the complex residual waveform is not well-filtered by a simple filter like $SG_6$, compared to the wavelet filter which is better matched to the shape of the transit signals of interest. On the other hand, precisely because it is simple and its 6 h bandpass matches the roll thruster firing period, $SG_6$ is a good metric for residual sawtooth. There may also be some stellar variability of the sample, since EPIC's photometric log g classifies ~50% of subgiants as dwarfs (Huber et al. 2015). We can minimize the impact of misidentified spectral types and stellar variability in general on measuring instrument performance with $CDPP_6$ and $SG_6$ by looking at 10$^{th}$ percentile noise, since quiet stars tend to be dwarfs. We can then more clearly attribute increased noise in *K2* to roll by noting that this 10$^{th}$ percentile noise per channel shows a clear increase with $r_{FOV}$ (Figure 5 and Table 1), exactly like the dominant source of error which present in *K2* but not present in *Kepler*. The increase of noise with $r_{FOV}$ is weak to nonexistent in *Kepler* data, at most 10% of that found in *K2*; instead, Gilliland et al. (2011) found instrument noise performance is dominated by focus, with the best CDPP on the annulus of best focus as can be seen in the left panel of Figure 5. While we suspect the star count $n_*$ may also be a factor in *K2* noise, as it is in compression (Section 2.3), we will not have *K2* Pipeline light curves on an FOV with the steep gradient in $n_*$ needed to measure this effect until C7 data processing is complete in May 2016. We thus fit each noise metric with only a constant term $a_0$ and a linear term $a_1$. See Perry (2015) for a preliminary discussion of *K2* photometry in crowded (high $n_*$) fields.

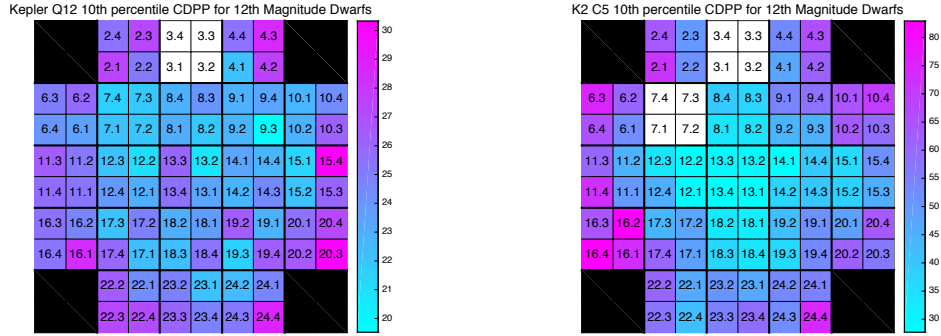

**Figure 5: 10th percentile 6 h CDPP for 12$^{th}$ magnitude dwarf (logg > 4) stars for *Kepler* (left) and *K2* (right), displayed as FOV images.** The corresponding FPA channels are labeled with module and output numbers. *Kepler* shows best performance on the annulus of best focus at intermediate radius and relatively little variation of CDPP across the focal plane compared to K2. Noise in *K2* increases linearly with radial distance from the center of the FOV, and hence is proportional to roll motion. Color maps are scaled to data ranges to show spatial patterns; note lower CDPP and CDDP variation in *Kepler* data.

**Table 1: Noise Metric Summary for 12th Magnitude Dwarf Stars.** The noise metrics in column 2 are ordered by time scale then campaign, and defined in the text. All noise metrics are 10$^{th}$ percentiles for dwarf (logg > 4) stars. <N> is the mean of the noise metric over all output channels, and $N_{cent}$ the average over the 4 central channels on module 13. $\sigma_N$ and $\sigma_{fit}$ are the standard deviation over the FOV before and after the linear fit. $a_0$ is the constant coefficient of the fit, and the $a_1$ the radial term in ppm/degree. $S_{a0}$ and $S_{a1}$ are the ratio of the coefficients to their standard errors; a value > 3 may be considered statistically significant. All *K2* noise metrics show significant radial dependence; a radial term is detectable in some *Kepler* data, but is an order of magnitude smaller. The full table, for magnitudes 11-15, is in the online version of this paper.

| Data Set | Noise Type | <N> | $N_{cent}$ | $\sigma_N$ | $\sigma_{fit}$ | $a_0$ | $a_1$ | $S_{a0}$ | $S_{a1}$ |
|---|---|---|---|---|---|---|---|---|---|
| *Kepler*-Q12 | $D_1$ | 61.9 | 60.7 | 4.2 | 3.5 | 55.5 | 1.3 | 48.4 | 5.9 |
| *K2*-C3 | $D_1$ | 109.2 | 65.8 | 28.6 | 16.2 | 45.1 | 13.0 | 8.3 | 12.6 |
| *K2*-C4 | $D_1$ | 102.3 | 65.7 | 24.7 | 12.3 | 44.1 | 11.9 | 10.7 | 15.0 |
| *K2*-C5 | $D_1$ | 87.1 | 69.1 | 16.4 | 10.7 | 53.2 | 6.9 | 14.8 | 10.0 |
| *K2*-C6 | $D_1$ | 93.3 | 68.2 | 18.5 | 10.4 | 51.6 | 8.5 | 14.8 | 12.7 |
| *Kepler*-Q12 | $CDPP_6$ | 24.8 | 24.8 | 2.0 | 1.8 | 21.9 | 0.6 | 38.0 | 5.3 |
| *K2*-C3 | $CDPP_6$ | 61.7 | 27.4 | 20.2 | 11.2 | 16.0 | 9.3 | 4.3 | 12.9 |
| *K2*-C4 | $CDPP_6$ | 61.9 | 26.6 | 19.6 | 10.5 | 16.7 | 9.2 | 4.7 | 13.7 |
| *K2*-C5 | $CDPP_6$ | 50.1 | 28.5 | 13.9 | 7.9 | 18.9 | 6.3 | 7.1 | 12.4 |
| *K2*-C6 | $CDPP_6$ | 53.3 | 29.1 | 14.7 | 8.3 | 20.4 | 6.7 | 7.3 | 12.5 |
| *Kepler*-Q12 | $SG_6$ | 25.9 | 25.7 | 2.2 | 1.9 | 23.1 | 0.6 | 36.8 | 4.8 |
| *K2*-C3 | $SG_6$ | 107.8 | 30.9 | 48.0 | 27.3 | 0.3 | 21.9 | 0.0 | 12.5 |
| *K2*-C4 | $SG_6$ | 105.3 | 31.4 | 45.4 | 25.5 | 3.1 | 20.8 | 0.4 | 12.7 |
| *K2*-C5 | $SG_6$ | 76.5 | 31.2 | 30.9 | 17.8 | 7.8 | 14.0 | 1.3 | 12.2 |
| *K2*-C6 | $SG_6$ | 83.2 | 32.2 | 31.1 | 17.6 | 13.5 | 14.2 | 2.3 | 12.5 |
| *Kepler*-Q12 | $SG_{72}$ | 30.0 | 20.2 | 7.3 | 6.3 | 19.6 | 2.1 | 9.5 | 5.3 |
| *K2*-C3 | $SG_{72}$ | 78.5 | 34.4 | 30.0 | 22.7 | 24.9 | 10.9 | 3.3 | 7.5 |
| *K2*-C4 | $SG_{72}$ | 68.9 | 24.2 | 28.5 | 20.6 | 15.3 | 10.9 | 2.2 | 8.2 |
| *K2*-C5 | $SG_{72}$ | 58.8 | 32.9 | 19.0 | 13.0 | 21.1 | 7.7 | 4.8 | 9.2 |

| | | | | | | | | |
|---|---|---|---|---|---|---|---|---|
| K2-C6 | $SG_{72}$ | 52.7 | 29.6 | 17.3 | 13.7 | 23.9 | 5.9 | 5.2 | 6.7 |

We find that $P_{10}CDPP_6$ averaged over all output channels increased from 25 to 50 ppm between *Kepler* and *K2*, while performance on the center module degraded by only 20%, from 25 to 30 ppm (Table 1). Figure 6 shows the $P_{10}CDPP_6$ as a function of $K_p$ for *Kepler* and *K2*, compared to noise models for *Kepler* and TESS (Sullivan et al., 2015); *K2* has lower noise than TESS for stars dimmer than 9th magnitude, though TESS compensates in part with a larger solid angle on the sky.  Figure 8 shows that while the FOV-averaged median per channel noise for dwarf stars has increased considerably between *Kepler* and *K2*, the noise shift for giants is relatively small so for these stars *K2* is operating close to astrophysical limits. The *K2* dwarf noise histogram also suggests a log-normal distribution of instrumental errors.

The C3 and C4 $CDPP_6$ was 20% higher than that of C5, which we attribute to apertures which were too small and therefore have more noise than necessary.  The problem was identified in the model-based TAD algorithm, which was not designed to cope with image motion exceeding one pixel, which led to poor inputs to subsequent steps. While TAD continues to serve well for defining target masks for data collection, of which the optimal aperture is a subset, it is no longer used help to define optimal apertures in collected data.  In C5 and later, only the data-driven approach (Section 3.2.2) is used.

The *K2* $P_{10}CDPP_6$ is also still significantly better than state-of-the art ground-based exoplanet photometry. Everett and Howell (2001) achieved 4.5 h noise close to the shot-noise limit using simple ensemble differential photometry, setting an upper bound to residual systematic noise sources of 170 ppm.  More recently, Nascimbeni *et al* (2013), using an RMS noise metric more like $SG_6$ than CDPP, achieved residual systematic noise of 150 ppm for ~2 hr transit duration, after RSS subtraction of the shot noise contribution.  The best *K2* $P_{10}CDPP_6$, from the central module, is far superior to that obtainable from the ground.  In addition, K2's continuous viewing gives it a decisive advantage for transits (or other phenomena) with durations exceeding 4 hr even in cases at the edge of the *K2* FOV where the photometric precisions are about the same as the very best ground-based photometry.

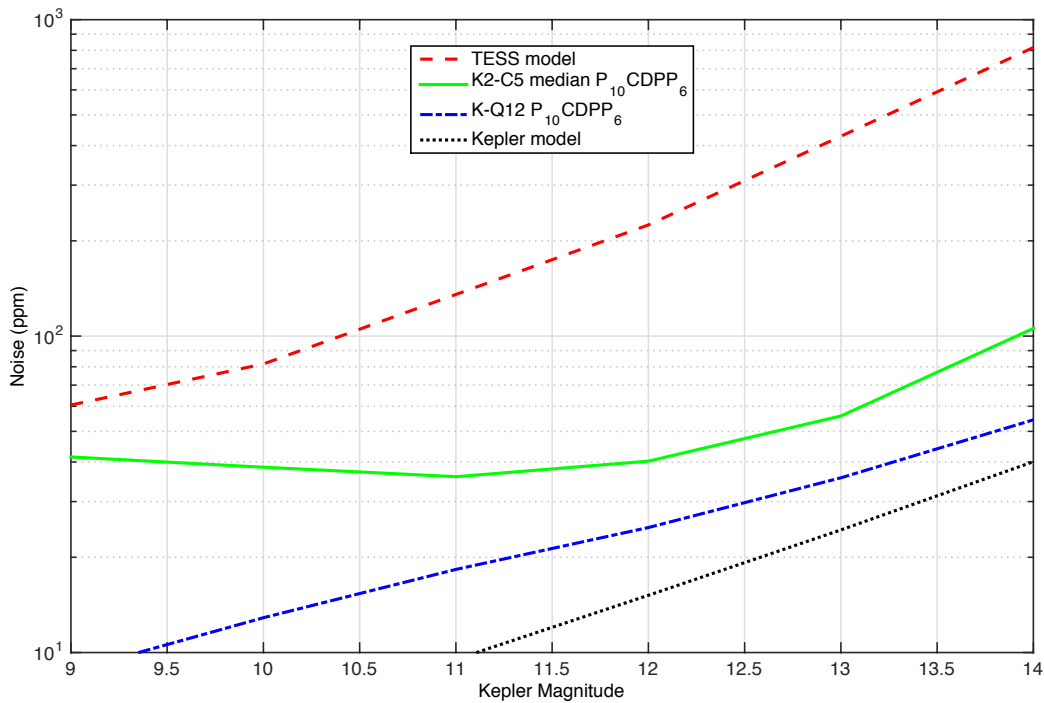

**Figure 6:  Comparison of Transit time scale noise for dwarf stars for *Kepler*, *K2,* and TESS.**  Model values include read, background, and source photocurrent shot noise, but do no include residual

systematic errors. The TESS model is from Sullivan et al. (2015), scaled to 6 h. See the online edition for a color version of this figure.

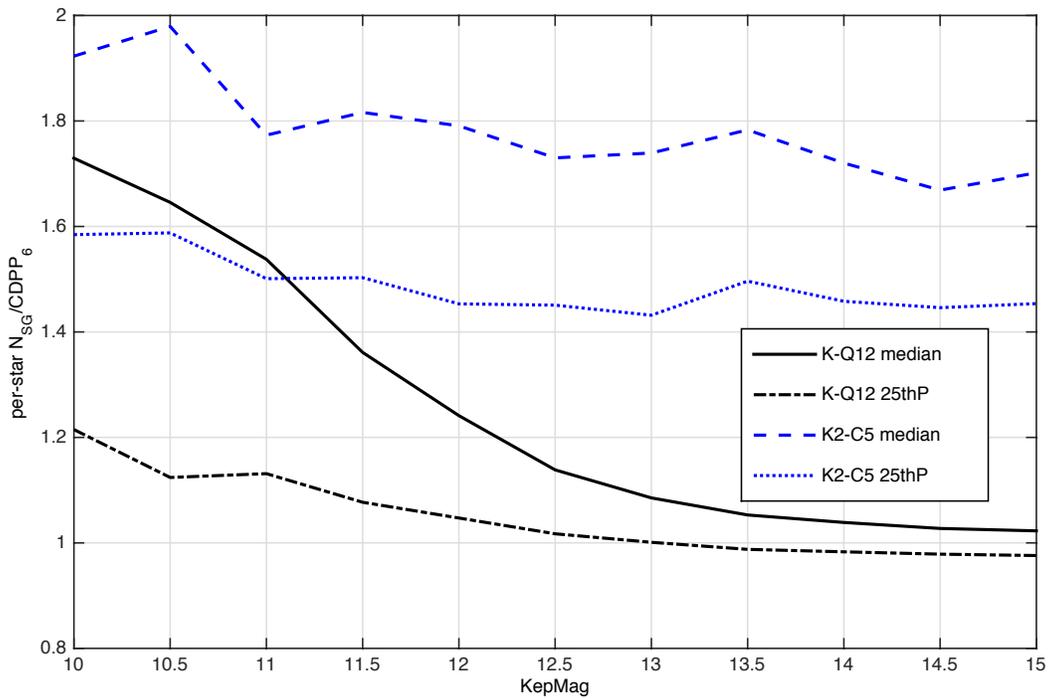

**Figure 7: Comparison of transit timescale noise metrics for dwarf stars.** Noise from a simple mid-frequency Savitsky-Golay filter $SG_6$ is a good proxy for percentile $CDPP_6$ for the *Kepler* mission, while the *K2* residual photometric errors on the same time scale make a simple filter like $SG_6$ less effective at isolating the noise corresponding to a transit detection process even in the quietest stars. Legend refers to percentiles of the ratio, not the ratio of percentiles. See the online edition for a color version of this figure.

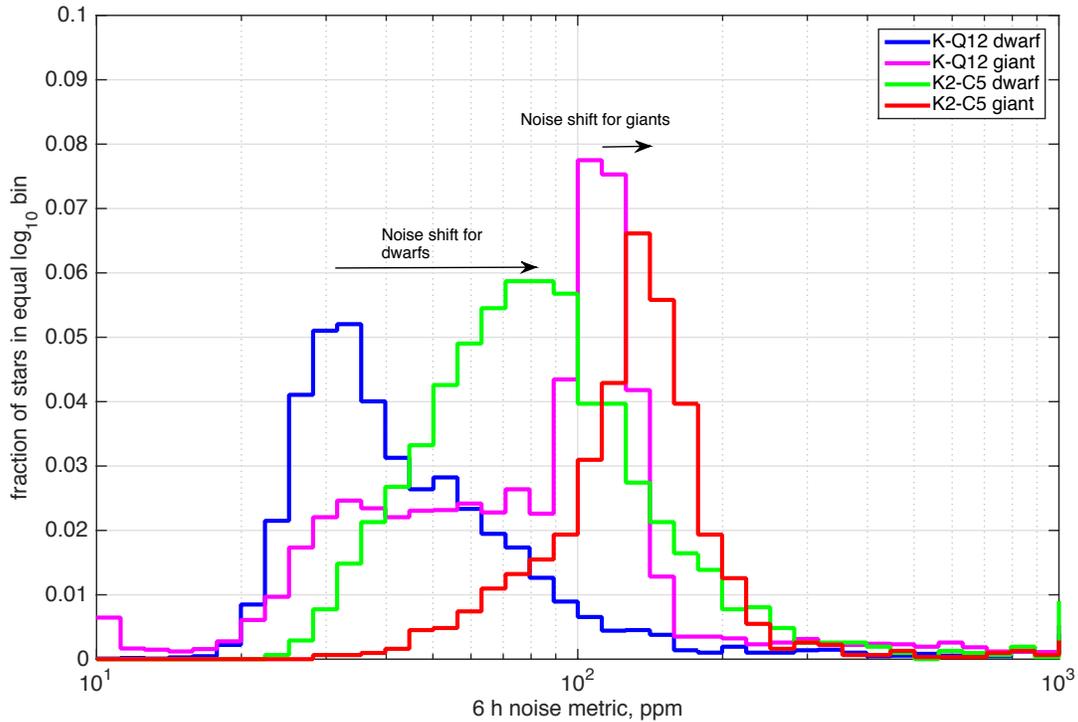

**Figure 8:  Comparison of full FOV CDPP histograms of *Kepler* Quarter 12 and *K2* Campaign 5 12$^{th}$ magnitude stars, identified as dwarf and giant stars from log g in the KIC and EPIC (Huber et al., 2015).**  *K2* remains nearly astrophysically limited for most giant stars.

### *4.3   Nyquist Noise*

While the 6 h timescale of CDPP$_6$ and SG$_6$ is a good match to the timescale of transiting exoplanets, the shortest timescales sampled by LC are relevant to asteroseismology.  Similar to the median (absolute) differential variability (MDV) of Lund et al. (2015) and Basri et al. (2013), we define a difference noise proxy D$_1$ for the Nyquist limit of LC data (corresponding to 2 LCs = 0.98 h = 283 µHz).  D$_1$ is the 5σ clipped standard deviation of the first temporal difference of normalized PDC-corrected flux value, divided by sqrt(2) to give unit result for zero-mean, unit variance WGN.  Since Fourier transforms of derivatives are equal to frequency times the transformed function, the filter corresponding to D$_1$ is equal to frequency from DC to the Nyquist limit. As suggested by Lund et al. (2015), we found a statistically significant radial increase in noise with distance from FOV center in *K2* data, as shown in Table 1.  As we found in Section 4.2 for SG$_6$, *K2* performance on the center module is within 20% of *Kepler* performance.  *Kepler* data also showed a weak radial trend in D$_1$, which was statistically significant but an order of magnitude less than that found for *K2*.

### *4.4   Long Timescale Noise*

Following Gilliland et al. (2015), we define a long time scale noise metric SG$_{72}$ which is long enough to be relevant to stellar rotation and variability signals, but not so long as to be influenced by the signal attenuation in the Pipeline for periods longer than 15 days (Section 6). SG$_{72}$ is calculated using the same procedure as SG$_6$, with a time scale 12 x longer (24 day quadratic polynomial and 3.0 day binning). As we found in Section 4.2 for SG$_6$, *K2* performance on the center module is within 20% of *Kepler* performance. *Kepler* data also showed a weak radial trend in SG$_{72}$, which was statistically significant but an order of magnitude less than that found for K2.

## 4.5 Correlations

The correlation between light curves give insight into the importance of systematic noise; stars which do not physically interact with each other are unlikely to be doing the same thing at the same time!  A simple measure of correlation is the Pearson correlation $C_{ij}$ between light curves *i* and *j*, which is the dot product of mean-subtracted, standard deviation-normalized light curves.  The "correlation goodness" metric used internally by Pipeline to decide which light curve algorithms give the best result for a particular target is (Stumpe et al. 2012):

$$G'_{C,i} = \frac{1}{\alpha_c \frac{1}{N_t} \sum_{j \neq i} |C_{ij}|^3 + 1} \qquad \text{Eq. 4-1}$$

where the functional form is chosen so that a "goodness" is confined to the interval [0,1] and a greater "goodness" means better data, $N_t$ is the number of targets, and $\alpha_c$ is a constant which determines the relative importance of correlation in the overall light curve correction assessment. $\alpha_c$ set equal to 12 for most of *Kepler* and for the *K2* mission to date.  The mean absolute correlation (MAC) $C_i$ for target *i* can be approximately recovered from existing Pipeline output by inverting Eq. 4-1:

$$C_i = \frac{1}{N_t} \sum_{j \neq i} |C_{ij}| \approx 0.686 \alpha_c^{-1/3} \left( \frac{1}{G'_{C,i}} - 1 \right)^{1/3} \qquad \text{Eq. 4-2}$$

where the numerical factor makes the equation an equality for WGN.  The RMS sum of $C_{ij}$ for WGN is $1/\sqrt{N_{samp}}$ where $N_{samp}$ is the number of samples, so $C_i$ less than this value are statistically insignificant.

While for noise statistics it is important to filter out astrophysically noisy stars, either by selecting dwarfs or using 10[th] percentile values or both, for correlation statistics we don't expect the astrophysical signatures of stars to be correlated with each other. So we compare the median (over all 12[th] magnitude stars on an output channel) MAC of K-Q12 and *K2*-C5 (Figure 9 and Table 2) and find that, just as with the noise statistics, the *K2* has a FOV-averaged MAC about twice that of *Kepler*, and a conspicuous radial dependence to which we fit a constant term $c_0$ and linear term $c_1$.  *Kepler* data show no statistically significant radial dependence.

Visual inspection of the *Kepler* data suggests that correlations are actually lower in regions of high star count, which may mean that there is a denser set of priors for MAP (Stumpe et al. 2012, Smith et al. 2012, Stumpe et al. 2014) to use in systematic error removal.  Because of the changes to PDC noted in Section 3.2.2, however, there is insufficient reason to believe *K2* will be like *Kepler*, and we will have to await light curves from campaigns of high star count and high star count gradient (C2 or C7) to look for this effect.

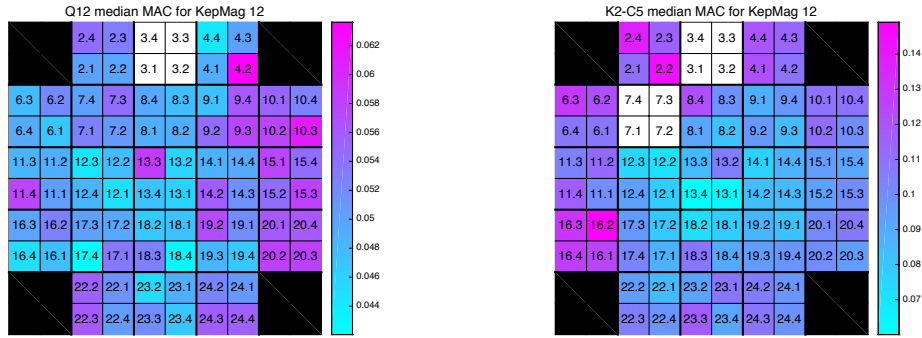

**Figure 9: Median mean absolute Pearson correlation of Pipeline-corrected light curves for 12th magnitude stars.** Residual correlations are twice as high in *K2*, and show an increase with radial distance from the center of the FOV as do the noise metrics.

**Table 2: Correlation Summary for 12th magnitude stars.** <C> is the FOV average of channel medians, and $C_{cent}$ the average over the 4 central channels on module 13. $\sigma_C$ and $\sigma_{fit}$ are the standard deviation over the FOV before and after the linear fit. $c_0$ is the constant coefficient of the fit, and the $c_1$ the radial term in ppm/degree. $S_{c0}$ and $S_{c1}$ are the ratio of the coefficients to their standard errors; a value > 3 may be considered statistically significant. All *K2* data show significant radial dependence.

| Data Set | <C> | $C_{cent}$ | $\sigma_C$ | $\sigma_{fit}$ | $c_0$ | $c_1$ | $S_{c0}$ | $S_{c1}$ |
|---|---|---|---|---|---|---|---|---|
| *Kepler*-Q12 | 0.051 | 0.049 | 0.005 | 0.004 | 0.048 | 0.001 | 33.6 | 2.8 |
| *K2*-C3 | 0.116 | 0.082 | 0.018 | 0.014 | 0.087 | 0.006 | 18.4 | 6.4 |
| *K2*-C4 | 0.098 | 0.077 | 0.020 | 0.015 | 0.063 | 0.007 | 12.2 | 7.1 |
| *K2*-C5 | 0.100 | 0.078 | 0.018 | 0.013 | 0.067 | 0.007 | 15.5 | 7.9 |
| *K2*-C6 | 0.091 | 0.060 | 0.015 | 0.010 | 0.062 | 0.006 | 19.0 | 9.8 |

## 5. Short Cadence Noise and Asteroseismology

Exquisite photometry is required in the milli-Hertz frequency regime in order to detect the small-amplitude, solar-like oscillations shown by cool main-sequence and subgiant stars.  The dominant oscillations have frequencies and corresponding amplitudes that range, respectively, from a few hundred micro-Hertz and 10 parts-per-million in subgiants close to the base of the red-giant branch; up to a few thousand micro-Hertz and a few parts-per-million in main-sequence stars like the Sun.  *Kepler* revolutionized the asteroseismic study of these solar-type stars (e.g., Chaplin et al. 2011), providing high-quality detections of oscillations in more than 700 targets, including more than 100 planet hosts. Short Cadence data, which extend Nyquist-sampled frequency coverage to 8500 µHz, have therefore been essential to *Kepler*'s asteroseismology work. The elevated levels of high-frequency noise expected for *K2* would inevitably present challenges for continuation of these studies.

Bright solar-type stars observed during the first few *K2* campaigns have provided an excellent test of the high-frequency performance of the new short-cadence data. The targets were chosen deliberately to sample a range of stellar parameters and apparent magnitudes that delivered detections during the nominal mission. Chaplin et al. (2015) recently presented the first results from these new *K2* short-cadence data. They employed the K2P$^2$ (*K2* Pixel Photometry; Lund et al. 2015) analysis pipeline and KASOC Filter (Handberg & Lund 2014) to extract and prepare lightcurves from archival calibrated pixels (Section 3.1 and 3.2.1) made in the first long campaign (C1). Analysis of the lightcurves revealed detections of oscillations in several subgiants. Chaplin et al. (2015) concluded that at C1-like levels of performance it would be possible to continue asteroseismic studies of cool subgiants showing dominant oscillation frequencies up to 1000 µHz, but not main-sequence stars. They anticipated that the increase in attitude control system bandwidth and consequent reduction in cross-boresight pointing error on SC timescales from C3 onwards (Peterson et al. 2015) would have the potential to deliver significant improvements, and suggested this might lead to detections of oscillations in main-sequence stars showing dominant frequencies as high as 2500 µHz. Here, we are able to report that the actual reductions in noise have exceeded expectations in C3 and C4, and that *K2* is now delivering levels of high-frequency noise that are typically within a factor of two (in power) of *Kepler*.

Figure 10 shows measures of the high-frequency power spectral density (robust average above 8000 µHz) in the frequency-power spectra of bright solar-type stars observed by *K2* (various campaigns, reduction by K2P$^2$) and *Kepler* (see legend). The dashed line follows the prediction given by the noise floor model presented in Gilliland et al. (2010) for the short-cadence *Kepler* data. The nominal-mission *Kepler* data (open circles) follow the noise model, as expected, where shot noise dominates the total noise. However, what is most striking about the figure is the improvement in the high-frequency *K2* noise from C3 onwards. Prior to this the total noise lay well above the expected shot-noise levels.

Figure 11 offers a window on the potential science return given by these improvements, showing clear detections of solar-like oscillations in solar-type stars observed in C3. The quality of the data is remarkable; the noise performance is achieved while preserving the signal. Particularly noteworthy are the two stars showing high-quality detections at frequencies above 3000 µHz (EPIC 206064678 and 206245055, in bottom row of figure). The results from C3 allow us to conclude that with *K2* it is now possible to perform asteroseismic studies of cool, bright main-sequence stars that have masses similar to the Sun.

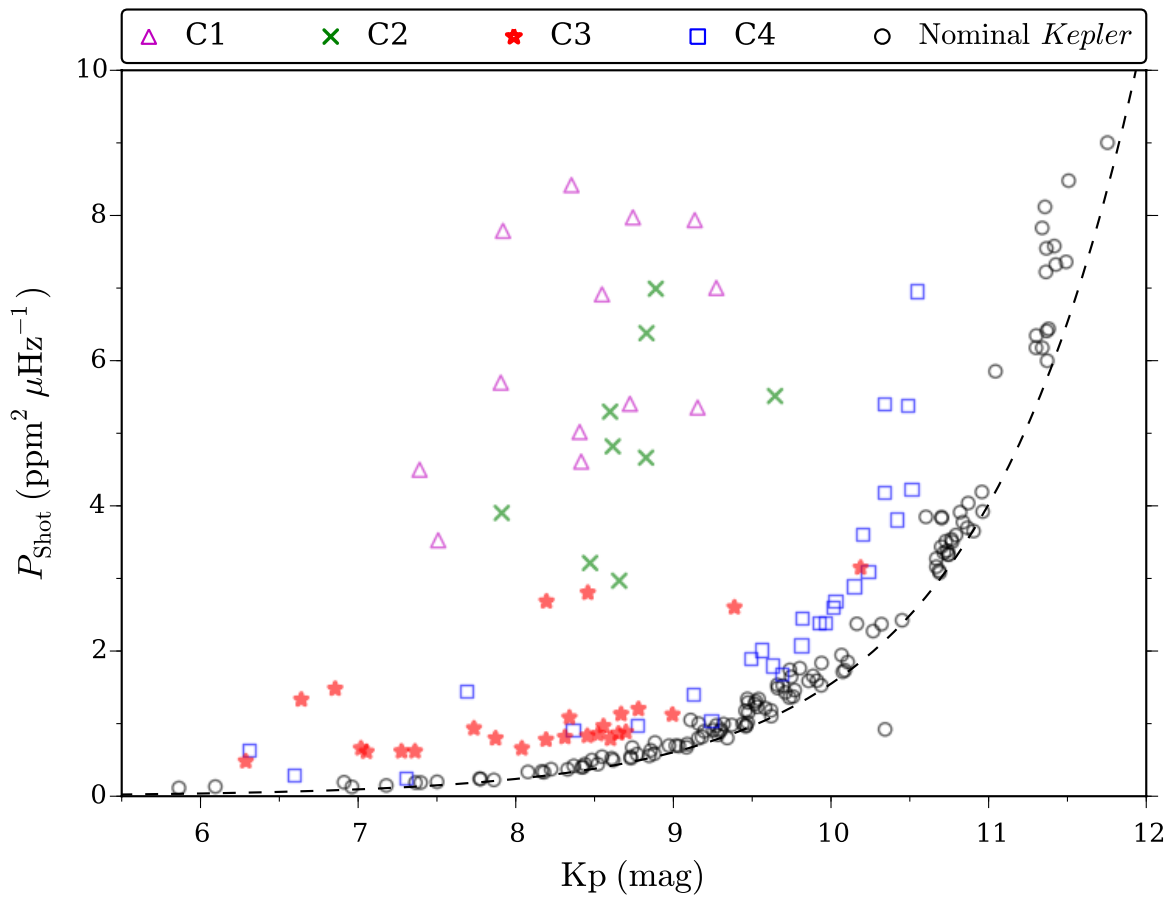

**Figure 10: Noise near the Short Cadence Nyquist limit (8500 μHz) for selections of solar-type stars observed by: *K2* in C1, C2, C3 and C4; and by the nominal *Kepler* Mission (see figure legend).** The dashed line follows the prediction of the SC noise model in Gilliland et al. (2010).

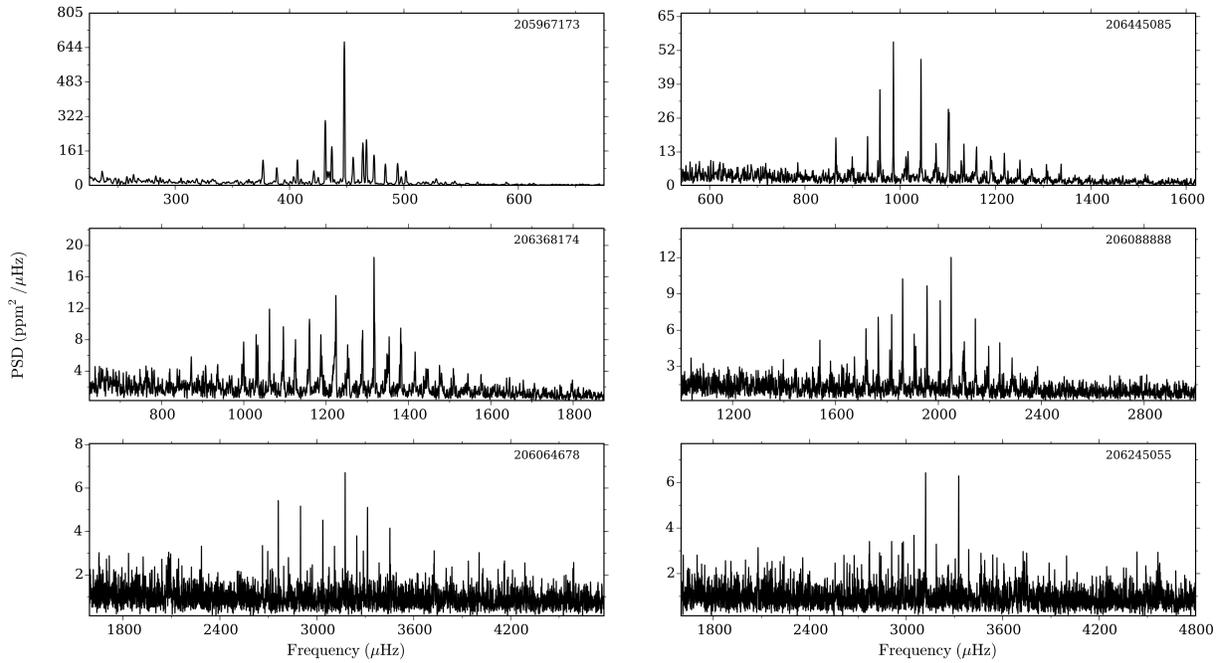

**Figure 11: Examples of *K2* SC frequency power spectra made from C3 data reduced by the K2P2 pipeline, showing clear detections of solar-like oscillations.**

## 6. Signal Fidelity

Good noise and correlation statistics do not necessarily mean good science; signal fidelity is equally important. Evidence for signal fidelity in *K2* archive light curves comes from the asteroseismology results for short cadence calibrated pixel data (Section 5), but also from sinusoid injection studies on Pipeline-corrected light curves.

Sinusoid injection studies were performed on C5 light curves to assure that removal of systematic errors did not remove astrophysical signals as well. Such injection studies were performed for *Kepler* data, as described in Gilliland et al. (2015). These studies involved injecting a single sinusoid onto the PDC input light curve for the 2081 targets on one of the 8 symmetrical FOV regions discussed on Section 3.2.2. The sinusoid amplitude was set to 1.0 times the standard deviation (after cubic detrending) of each underlying SAP light curve. No cuts were made on the targets in the study (e.g. magnitude, logg, etc…) which means a realistic distribution of target light curves was represented in the study. The period distribution of the sine waves is as given in Figure 12 and with a random phase distribution. The PDC output light curves are then compared to a reference run with no injected sine-waves. The preservation of the injected sine waves is then characterized as an amplitude attenuation as shown in Figure 12. Signals less then 15 d period are not significantly attenuated, while signal loss increases rapidly beyond that. Hence the $SG_{72}$ metric, which has an effective bandpass of 8 to 15 days, is at the upper limit of Pipeline signal fidelity.

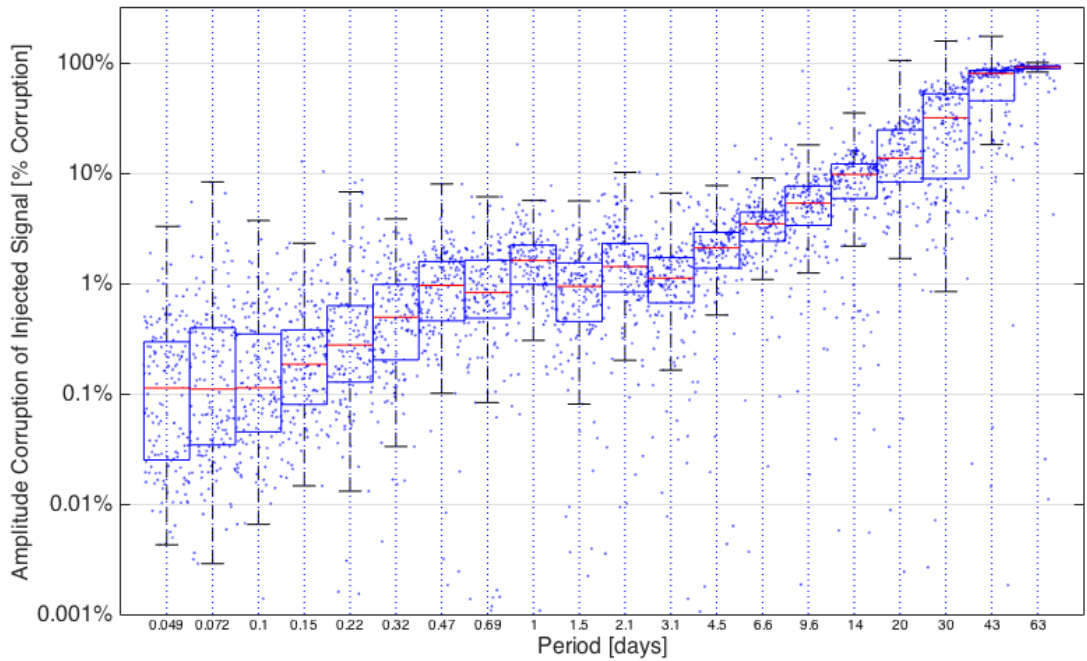

**Figure 12: Signal loss for sinusoids in *K2* corrected light curves**, showing < 10% signal loss for periods up to 15 days, in agreement with the *Kepler* results shown by Gilliland et al. (2015). The amplitude of the injected signal is equal to the standard deviation of the cubic-detrended uncorrected SAP light curve. For each period bin, the horizontal red line inside the box shows the median, box limits are at 25% and 75%, and the whiskers are at approximately +/–2.7σ. See the online edition for a color version of this figure.

## 7. Summary and Conclusions

In the first 18 months of *K2,* we've learned how to increase the total number of targets observed and reduce the average noise by improvements in how we compress the data, point the telescope, allocate pixels, and analyze the light curves. This paper shows that the compression, noise, and residual correlations in archive light curves depend on radial distance from the focal plane center, and that compression also depends on star count.  The compression results may be useful in getting "more bang for the bit" out of K2's storage and downlink budget in an extended *K2* mission, while the noise results will give data users an estimate for residual systematic noise in archival light curves for a given campaign and channel.  These noise results will be helpful for future GOs, who may wish to select targets near the center of the FOV for best noise performance.

Funding for this Discovery Mission is provided by NASA's Science Mission Directorate. We thank the Kepler Science Operation Center and Science Office staff whose efforts led to the data products discussed in this work. We thank in particular Tom Barclay and Fergal Mullally for reading early drafts and making helpful comments; Wendy Stenzel for Figure 1; Mike Haas and Charlie Sobeck for K2 project support of the preparation of this manuscript; Daniel Huber for the EPIC stellar properties in advance of publication; and Ball Aerospace and LASP for making the operational improvements which led to these results.  This work was supported by NASA grant NNX13AD01A.